\documentclass[manuscript,screen,nonacm]{acmart}

\AtBeginDocument{%
  }

\graphicspath{{./images/}}
\PassOptionsToPackage{hyphens}{url}
\usepackage{hyperref}
\hypersetup{colorlinks=true, linkcolor=blue, urlcolor=blue, citecolor=blue, breaklinks=true}
\usepackage{xurl}
\usepackage{booktabs}
\usepackage{array}
\usepackage{xcolor}
\newcolumntype{L}[1]{>{\raggedright\arraybackslash}p{#1}}
\makeatletter
\newenvironment{appitems}{\enumerate\def\p@enumi{\thesubsection.}}{\endenumerate}
\makeatother

\usepackage{tabularx}
\newcolumntype{Y}{>{\raggedright\arraybackslash}X}
\usepackage{arydshln}
\usepackage{needspace}
\newcommand{\rowsep}{\hdashline[0.4pt/1.5pt]\noalign{\vskip 2pt}}

\newcommand{\br}{\discretionary{}{}{}}
\begin{document}

\title{A Code-Level Audit of X}

\author{Luka Bekavac}
\email{luka.bekavac@unisg.ch}
\orcid{0009-0009-3598-3012}
\affiliation{%
  \institution{University of St. Gallen}
  \city{St.Gallen}
  \country{Switzerland}
}

\author{Simon Mayer}
\email{simon.mayer@unisg.ch}
\orcid{0000-0001-6367-3454}
\affiliation{%
  \institution{University of St. Gallen}
  \city{St.Gallen}
  \country{Switzerland}
}

\renewcommand{\shortauthors}{Bekavac and Mayer}

\begin{abstract}
The transparency regime of recommender systems on Very Large Online Platforms (VLOPs) is undergoing significant change under the EU Digital Services Act (DSA). The DSA requires platforms to disclose information about their recommendation pipelines and grants vetted researchers access to non-public data, which may extend to recommender code. However, it is not known \emph{to what extent} such code releases enable \emph{substantive and meaningful} third-party auditing of a production recommender system. We address this question by proposing a code-level audit methodology for recommender systems and applying it to the X platform, which has voluntarily open-sourced its recommender pipeline. Our methodology is a backwards-walk traversal of the repository, tracing each decision-relevant component from the user-facing timeline back to its upstream producers, checking that it is reachable from the served pipeline, and classifying each under a four-tier inspectability scheme. Applying this procedure, we identify \emph{13 transparency dead-ends} on the live serving path: models, services, label lists, and configuration values that the public code depends on but the release does not contain. We find that the components that decide which posts are filtered and how they are ranked fall into the two least inspectable tiers, a pattern that intensifies with the 2025 integration of the proprietary Grok language model into moderation, annotation, and ranking. We further show that the release keeps 2023 code that the 2025 pipeline no longer reaches, with 36 parameter references that would pass for dead-ends without a reachability check. Finally, we discuss which additional artifacts would be required for meaningful auditing.
\end{abstract}



\maketitle
\section{Introduction}

Platforms such as X shape public discourse through large-scale recommender systems that determine which posts users see, in what order, and with what prominence. These systems integrate personalization, ranking, and moderation into pipelines that optimize for relevance and for platform-specific economic and political objectives~\cite{Leerssen_soap}. Their outputs co-determine exposure to disinformation, discriminatory content, and other systemic risks, making them a central regulatory concern~\cite{Messmer_Degeling_2023}.
Despite this scrutiny, recommender systems remain opaque, owing to both technical complexity and deliberate corporate secrecy (cf.~\cite{opacity-ML}). Most empirical work therefore infers algorithmic behavior from outputs rather than from internal design and relies on so-called black-box methods---sock-puppet audits, scraping, network analyses, and observational studies---which themselves face methodological, ethical, and legal constraints~\cite{Weird-Audits, bandy2021problematicmachinebehaviorsystematic, Social_Data_Pittfalls, TORS}. For AI audits, researchers have argued more generally that black-box access is insufficient and that auditors also need white-box and ``outside-the-box'' access, for instance to code, documentation, and deployment details~\cite{casper2024blackbox}. In addition to only providing an incomplete view of the external behavior of a system, these methods can merely show \emph{that} a system behaves in a particular way, but not \emph{why}. On the other hand, code-level audits---which could address this gap---barely exist as a genre in the domain of recommender systems: a 2025 systematic review of 176 algorithm-auditing studies found only a single study that inspected source code directly, and none targeted a recommender system~\cite{Weird-Audits, mobile-audit}.

The EU Digital Services Act (DSA)~\cite{DSA2022} addresses these concerns. It requires very large online platforms (VLOPs), services with more than 45 million monthly users in the EU, to assess and mitigate the \emph{systemic risks} their services create, such as the spread of illegal content and negative effects on fundamental rights, civic discourse, electoral processes, public health, and minors (Arts.~34 and~35). The design of recommender systems is named explicitly as a factor in these risks. The DSA pairs these duties with transparency obligations that allow users, regulators, and researchers to scrutinize how recommender systems work: Art.~27 requires platforms to disclose their main recommender parameters in plain language; Art.~38 requires VLOPs to offer at least one recommender option not based on profiling; and Art.~40(4) grants vetted researchers access to non-public data necessary to assess systemic risks, under formal access controls~\cite{Seiling_Clara_Ohme_Klinger_Claes_2025}. While nothing in Art.~40(4) excludes it, the DSA does not name source code, and regulators can already demand that VLOPs explain the design and logic of their recommender systems (Art.~40(3)). Platforms typically satisfy DSA transparency obligations through structured documentation, audit reports, and negotiated researcher access. As data-access requests under the DSA become more common, independent researchers will have increasingly the opportunity to scrutinize source code directly and code-level audit methodologies will matter well beyond the case of voluntary releases.
X has voluntarily open-sourced substantial parts of its production recommender pipeline, first in 2023\footnote{Twitter Engineering, \href{https://blog.twitter.com/engineering/en_us/topics/open-source/2023/twitter-recommendation-algorithm}{Twitter's Recommendation Algorithm}, 31 March 2023.} and again in a major 2025 update. X made this release voluntarily, outside any DSA obligation, and framed it at the time of the 2023 release as enabling independent third parties to ``determine, with reasonable accuracy, what will probably be shown to users'' (App.~\ref{app:tweet-opensource}). Within this regulatory landscape, the present study treats X's release as a test case: because source code is provided, gaps between regulatory transparency requirements and platform disclosures become analytically visible in our study.

The 2025 update is also architecturally distinctive: It extends the pipeline with X's proprietary large language model (LLM), \emph{Grok}, into moderation, annotation, and ranking logic. This is a transition to a partially non-deterministic hybrid Code/LLM recommender architecture whose semantic annotations, safety classifications, and contextual embeddings now co-determine what content users see. This shift creates a new form of transparency gap: While the deterministic components remain visible in the open-sourced code, the language model layer---its prompts, weights, and inference logic---remains hidden and/or closed. It is therefore an open question which parts of the 2025 pipeline remain auditable, and how LLM integration reshapes the auditability of production recommender systems across platforms more generally. The broader industry is moving in the same direction, which increases the relevance and urgency of this question. Meta's CEO Mark Zuckerberg has stated in the company's Q4~2025 earnings call that Meta is ``working on merging LLMs with the recommendation systems that power Facebook, Instagram, Threads, and our ads system.''\footnote{Meta Platforms Q4~2025 earnings call \url{https://www.facebook.com/zuck/posts/pfbid0XaqxVGXPkGHRnotatVneydxapZsSjHbBXVPuDnJGC9T6P3Z76MeenyiW6H3c69asl}.}
As other VLOPs follow, the auditability questions raised by X's Grok integration will apply across the sector, and the European Commission has already opened a formal investigation into risks associated with Grok's deployment in the EU.\footnote{\url{https://ec.europa.eu/commission/presscorner/detail/en/ip_26_203}}

With this paper, we provide methodological and empirical contributions to close this research gap. Specifically:
\begin{enumerate}
\item We propose a \textbf{generalizable methodology} for code audits of VLOP recommender systems, comprising a backwards-walk traversal procedure with a reachability check, a four-tier inspectability classification, and a systematic dead-ends inventory.
\item We \textbf{apply this methodology to X's 2025 open-source repository}, documenting how user segmentation, visibility filtering, and content ranking are implemented, and how Grok-mediated annotation signals reshape the recommendation pipeline.
\item We present a \textbf{dead-ends inventory} of 13 transparency gaps: models, services, label lists, and configuration values that the public code depends on but the release does not contain. For each, we show which audit questions the gap forecloses.
\end{enumerate}

\section{Background and Related Work}

Existing audit methodologies cluster around two access regimes. The first is first-party access, where organizations audit their own systems with full internal access to design documents, engineers, and configuration. The second is third-party observational access, where researchers study system outputs without any access to internal design. The case this paper addresses falls between these two regimes: Under data access through the DSA or voluntary disclosure, third parties gain access to partial source code but not to engineers, weights, or production configuration. 
In the following, Secs.~\ref{sec:audit-methods}--\ref{sec:systemic-risk} review audit methods and frameworks, Sec.~\ref{sec:code-audit-gap} the few existing code audits, and Secs.~\ref{sec:reproducibility-limits} and~\ref{sec:provenance} the limits of code releases and work on decision provenance.

\subsection{Audit Methods for Platform Recommender Systems}
\label{sec:audit-methods}

Sandvig et al.~\cite{sandvig2014auditing} first distinguished research designs for auditing platform algorithms; building on their taxonomy, Bandy~\cite{bandy2021problematicmachinebehaviorsystematic} identifies five methods: code audits, direct scrapes, sock-puppet audits, carrier-puppet audits, and crowdsourced audits.
The vast majority of empirical work falls into the latter four categories, inferring algorithmic behavior from outputs rather than from internals.
This hence yields valuable insights into observed effects but provides limited understanding of underlying mechanisms, and faces methodological~\cite{Mosnar_2025}, ethical~\cite{Social_Data_Pittfalls}, and legal~\cite{TORS} constraints.

\subsection{First-Party and Internal Audit Frameworks}
\label{sec:internal-audits}

A separate line of work proposes audit frameworks for organizations auditing their \textit{own} systems before deployment. Raji et al.~\cite{raji2020closing} present SMACTR, an end-to-end internal audit framework with five documented stages: \textit{Scoping}, \textit{Mapping}, \textit{Artifact Collection}, \textit{Testing}, and \textit{Reflection}.
The framework assumes \emph{cooperative access} to design documents, engineers, and risk-assessment processes inside the developing organization.
The methodology we propose audits from outside the organization, with only a partial codebase available, and therefore complements SMACTR. The dead-ends inventory we produce in Sec.~\ref{sec:dead-ends} corresponds to what the \textit{Mapping} and \textit{Artifact Collection} stages of SMACTR yield when access to engineers, stakeholders, and undisclosed configuration is foreclosed.

\subsection{Systemic-Risk Audit Frameworks}
\label{sec:systemic-risk}

A second strand of work frames recommender auditing around systemic risks at the VLOP level. Meßmer and Degeling~\cite{Messmer_Degeling_2023} propose a risk-scenario-based audit methodology that translates systemic-risk categories into actionable hypotheses and measurement strategies, warning that the absence of concrete methodologies risks promoting ``audit-washing.'' Leerssen~\cite{Leerssen_soap} maps the European transparency landscape to three categories of disclosure: user-facing notices, government auditing, and data-sharing partnerships, and argues that a fourth---\textit{public data access}---is needed for inclusive scrutiny.
Panigutti et al.~\cite{Delia-FAccT} provide a taxonomy of algorithm audit study designs that structures what questions an audit should ask.
This literature assumes observational or quasi-experimental access only, and there is no methodology for navigating a public or otherwise accessible production codebase.

\subsection{Code Audits of Production Recommender Systems}
\label{sec:code-audit-gap}

Code audits remain rare: the only one identified in the systematic review discussed in the introduction~\cite{Weird-Audits}, by Weber and Kosterich~\cite{mobile-audit}, analyzed 59 open-source mobile news applications and demonstrated how editorial and normative decisions become embedded in software through filtering, sorting, and ranking routines. %

Our study is based on a rare opportunity: X's 2023 release published substantial parts of the recommender-system codebase under an open-source license, the first VLOP to do so at this level of detail.%
Shortly after the 2023 release, Messing documented in a widely-read blog post\footnote{S.~Messing, \href{https://solomonmg.github.io/blog/twitter-the-algorithm/}{What can we learn from ``The Algorithm,'' Twitter's partial open-sourcing of its feed-ranking recommendation system?}, April 2023.} that the code was there, but missing specifics. Key parameters, feature sets, and model weights were absent or abstracted. Messing reported engagement weights from a README in the companion ML repository, under which a reply counted 27 times as much as a like and a reply that the post's author engaged with 150 times as much, but cautioned that the release alone could not explain the algorithm without the underlying model features, parameters, and data. Subsequent commentary on the 2025 update consists largely of informal blog posts and practitioner notes that aim to ``reverse engineer'' the algorithm or optimize content for broader reach.\footnote{E.g.\ \url{https://medium.com/@byteHumi/i-analyzed-twitters-source-code-here-s-how-to-actually-get-a-banger-no-bs-da4f724ad8c2}} These resources draw on README files or isolated code snippets and lack systematic analytical grounding. They surface individual nuggets but do not induce a framework for navigating the codebase as a whole, and the 2025 Grok integration has not been analyzed in the scholarly literature at all.

\subsection{Limits of Reproducibility in Code Releases}
\label{sec:reproducibility-limits}

Releasing code does not guarantee reproducibility, even in academic recommender systems where less competitive pressure applies. Hidasi and Czapp~\cite{hidasi2023thirdparty} demonstrated that results shift materially depending on which third-party implementation of a baseline is used. In a companion paper, they showed that widespread flaws in offline evaluation setups propagate across the literature as later works copy earlier evaluation protocols without scrutiny~\cite{hidasi2023widespread}.
~\cite{ferrari2021troubling} found that 11 of the 12 neural recommendation approaches they could reproduce, all published at prestigious conferences between 2015 and 2018, were outperformed by conceptually simple baselines. Releasing source code is hence necessary, but not sufficient, for reproducibility or auditability: source code alone rarely shows how a decision was made, because its behavior depends on inputs and operating conditions that the code does not contain~\cite{kroll2017accountable}; in addition, outsiders cannot confirm that disclosed code is the code that runs~\cite{kroll2017accountable}. The pattern recurs in the VLOP setting we examine, where dependencies, configurations, and tuning procedures that determine production behavior are routinely absent even when code is shared.

\subsection{Decision Provenance and Traceability}
\label{sec:provenance}

A separate research tradition argues that algorithmic accountability requires tracing the chain of inputs to, the nature of, and the flow-on effects from decisions across interconnected components. Singh, Cobbe, and Norval~\cite{singh2019decision} call this \emph{decision provenance} and position it against narrow algorithm-focused approaches: accountability requires data-flow exposure across technical and organizational boundaries, not just model inspection. Mora-Cantallops et al.~\cite{moracantallops2021traceability} review traceability tools for AI systems, including W3C PROV-based approaches, and define minimal requirements for a model to be traceable.
Within recommender systems specifically, user-centric evaluation research treats transparency, i.e., whether users understand why an item was recommended, as a core quality of the system~\cite{pu2012traceability}, but it addresses what users perceive rather than where the underlying information comes from. The literature, hence, articulates what should be traceable or transparent in algorithmic systems; but it does not empirically maps where provenance actually breaks down in a production VLOP recommender. 

The methodology we propose closes this gap. Our proposed \textit{four-tier inspectability classification} follows the idea of a provenance-status classification, and our proposed \textit{dead-ends inventory} is an empirical map of where provenance breaks down in a real VLOP codebase.

\section{Method}
\label{sec:methodology}

A code-level audit of a production recommender system has long been close to infeasible in practice. X's open-sourced repository spans roughly 615{,}000 lines of code, mostly Scala and Java, spread over dozens of services that exchange data through shared feature definitions and remote calls. Reading it bottom-up is not a realistic method, and even a complete reading would miss most of what drives user-visible behavior---weights, thresholds, experiment configurations, LLM prompts---because these are not part of the release. Our methodology is therefore \textit{top-down}: We start from what the platform itself states about the system and use those statements to decide which components to inspect. The primary analytical output is an inventory of \textit{dead-ends}: points at which code that actually runs depends on a parameter, service, model, or data source that the release does not contain.

\subsection{Audit Inputs}
\label{sec:audit-inputs}

The audit draws on sources with different evidentiary status. \emph{Secondary} sources tell us what the platform claims and which components it describes as new: X's Transparency Center\footnote{\url{https://transparency.x.com/en}} and Help Portal,\footnote{\url{https://help.x.com/en/resources/recommender-systems}} and posts from verified X corporate, engineering, and leadership accounts (App.~\ref{app:xaccounts}). We use them only to decide what to inspect. \emph{Primary} sources are used to verify or falsify those claims: the repository at commit \texttt{c54bec0},\footnote{\url{https://github.com/twitter/the-algorithm/tree/c54bec0d4e029fe34926ef3258a86ccacc0d0182}} its commit history, and its inline documentation (e.g., App.~\ref{app:trust-safety-readme}). The 2025 update arrived as a single commit adding 65{,}319 and removing 3{,}195 lines, so the diff localizes what changed, including the Grok-related additions. No code claim in this paper rests on transparency materials alone.

\subsection{Backwards-Walk Audit Traversal}
\label{sec:traversal}

A codebase of this size needs a systematic way of deciding what to read and in which order. We call our procedure a \emph{backwards walk}. It starts where recommendations become visible to users---the assembled timeline---and follows each decision back through the code to the components and data that produce it, much as one would trace a river upstream to its sources. Walking backwards keeps the audit focused on code that actually shapes what users see, and it ends where the trail leaves the release: at the dead-ends we want to document. The traversal consists of five steps:

\begin{enumerate}
    \item \textbf{Flag what to inspect.} From the secondary sources, list the components described as live or recently changed---here, the Grok integration, verification tiers, Communities, evergreen content, and the trust-and-safety classifiers.
    \item \textbf{Enter the code at the user-facing surface.} Start from the code that assembles the timeline and follow its wiring to each flagged component (Sec.~\ref{sec:pipeline-overview} explains the building blocks).
    \item \textbf{Check reachability.} A release can contain code that nothing calls. Keep only components that are reachable from the pipeline actually served, and set the rest aside instead of classifying it.
    \item \textbf{Trace each component to its producers.} For every signal a component uses, follow it upstream until it either resolves inside the repository or ends in an external service, an undisclosed module, or a configuration value set outside the code.
    \item \textbf{Classify and record.} Assign each component an inspectability class (Sec.~\ref{sec:inspectability}), and record every unresolved dependency as a dead-end.
\end{enumerate}

This procedure produces a directed graph of dependencies with dead-ends as leaf nodes.

\subsection{Inspectability Classification}
\label{sec:inspectability}

We define four tiers of inspectability for components of a hybrid Code/LLM recommender system:

\begin{description}
    \item[Class~I (Fully inspectable):] Both the logic and its parameters are present in the repository, so the component's behavior can be reconstructed from the code alone. \emph{Example:} the formula that scales X's reputation score to a range of 0--100 (App.~\ref{app:tweepcred}).
    \item[Class~II (Parameter-dependent):] The logic is visible, but the values it runs with are set outside the code. The auditor can identify \emph{what} the code does but not \emph{how much} it does in production. \emph{Example:} the factor that down-weights posts from accounts the viewer does not follow, whose code default of 0.75 production may override (App.~\ref{app:oon-scale}).
    \item[Class~III (Interface only):] The call is visible, but the implementation resides in an external service or proprietary model. The auditor can identify \emph{that} a decision is made and \emph{which signals it produces}, but not \emph{how}. \emph{Example:} the component that attaches Grok's annotations to each post (App.~\ref{app:grok-hydrator}).
    \item[Class~IV (Opaque):] The component is referenced by name only, with no interface, implementation, or documentation in the repository. \emph{Example:} the internal table that lists ``evergreen'' content (App.~\ref{app:evergreen-label}).
\end{description}

\section{Application: Auditing X's Codebase}
\label{sec:worked-example}
We apply the methodology to X's 2025 open-source repository. We first give a general overview of how X's recommender system works (Sec.~\ref{sec:pipeline-overview}) and then follow a single post through it (Sec.~\ref{sec:trace-steps}); Secs.~\ref{sec:findings} and~\ref{sec:dead-ends} report what the traversal surfaced. All code references are pinned to commit \texttt{c54bec0}, the most recent commit at the time of analysis, which is permanently archived in Software Heritage (App.~\ref{app:source-code-refs}). Every file path and line reference is verified against this commit and listed in the appendix as a GitHub permalink.

\subsection{Architecture of X's Recommender}
\label{sec:pipeline-overview}

\begin{figure*}[t]
    \centering
    \includegraphics[width=\textwidth]{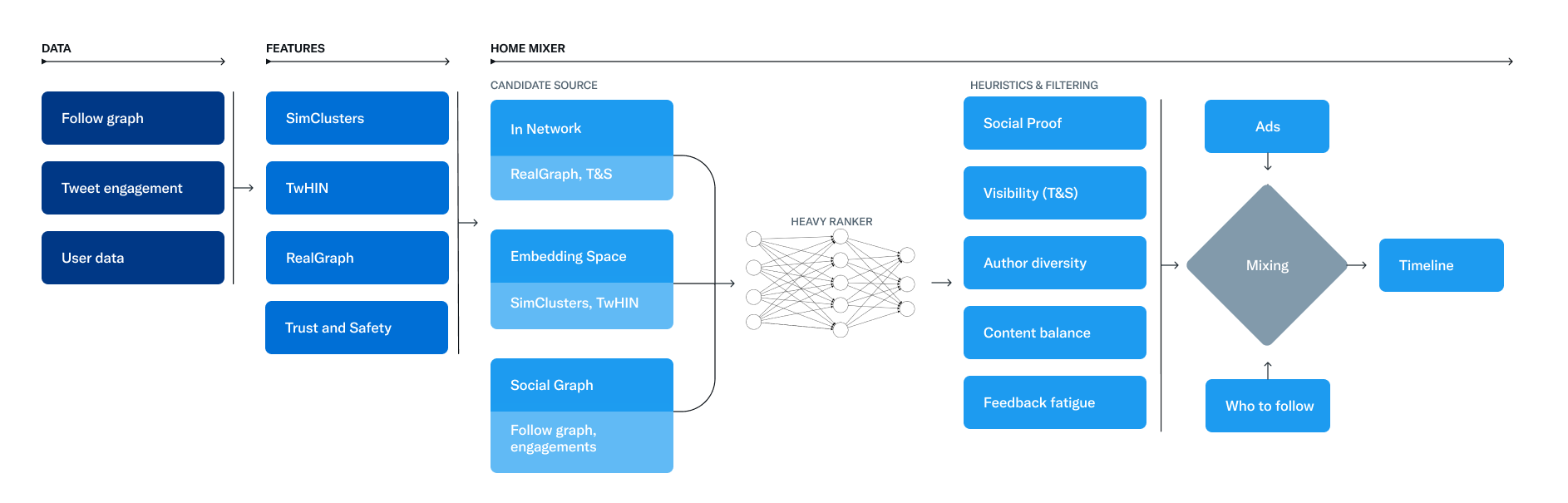}
    \caption{High-level architecture of X's Home Timeline recommendation pipeline. Source: \url{https://github.com/twitter/the-algorithm}.}
    \label{fig:x-architecture}
    \Description{Left-to-right flow diagram of X's Home Timeline pipeline. Three data sources (follow graph, tweet engagement, user data) feed four feature components: SimClusters, TwHIN, RealGraph, and Trust and Safety. These feed the Home Mixer, which first draws candidates from three sources: In Network (RealGraph, T\&S), Embedding Space (SimClusters, TwHIN), and Social Graph (follow graph, engagements). The candidates converge on a neural-network icon labeled Heavy Ranker. The ranked posts then pass through heuristics and filters for social proof, visibility (T\&S), author diversity, content balance, and feedback fatigue. A diamond labeled Mixing combines the filtered posts with Ads and Who to follow modules into the Timeline.}
\end{figure*}

When a user opens the \emph{For You} timeline, X's \texttt{home-mixer} service assembles it in four stages (Figure~\ref{fig:x-architecture}). \emph{Candidate sourcing} gathers some 1,700 posts: \emph{in-network} posts from accounts the user follows and \emph{out-of-network} posts from accounts the user does not follow, which are found through models of users' interests and interactions. \emph{Ranking} scores each candidate with machine-learning models that predict how the user will engage with it. \emph{Filtering}, both before and after ranking, removes posts that are too old, already seen, or flagged as unsafe. \emph{Mixing} assembles the final timeline and interleaves the ranked posts with advertisements and recommendation modules. Table~\ref{tab:glossary} introduces the building blocks the code uses for these stages; we refer to them throughout the paper.

\begin{table}[!ht]
\centering
\small
\caption{Building blocks of X's recommendation code}
\label{tab:glossary}
\begin{tabular}{L{0.18\textwidth} L{0.5\textwidth} L{0.24\textwidth}}
\toprule
\textbf{Building block} & \textbf{What it does} & \textbf{Example in X's code} \\
\midrule
Pipeline & A configured sequence of steps that produces part of the timeline. Pipelines are nested, and only those the product registers actually run. & \texttt{ForYouMixer\-PipelineConfig} \\
Candidate source & Retrieves posts that might be shown, from a search index, an interaction graph, or a similarity model. & \texttt{tweet-mixer}, SimClusters \\
Feature & A piece of information about a post, its author, or the viewer, such as the author's verification status or a post's predicted topic. & \texttt{AuthorIsBlue\-VerifiedFeature} \\
Feature hydrator & Fetches features from other services and attaches them to each candidate---it ``hydrates'' a bare post ID with information---so that later steps can use them. & \texttt{GrokAnnotations\-FeatureHydrator} \\
Filter & Removes candidates that fail a condition, such as being older than 48 hours. & \texttt{GrokSpamFilter} \\
Ranking model and heads & Predicts, for each candidate, the probability of each kind of engagement (like, reply, repost, \ldots); each prediction is a \emph{head}. A weighted sum of the heads is the candidate's score. & Navi (default), Phoenix \\
Rescoring factor & Multiplies the score after the model, e.g., to down-weight out-of-network posts or to diversify authors. & \texttt{OutOfNetwork\-ScaleFactorParam} \\
Parameter (feature switch) & A named configuration value. The code ships a default; production values are set at runtime and can differ between experiment groups. & \texttt{EnableGrok\-Annotations} (default off) \\
External service & A system the code calls but the release does not contain, such as the user database or a model server. & Gizmoduck (users), Navi (model) \\
\bottomrule
\end{tabular}
\end{table}

\subsection{End-to-End Trace of a Candidate}
\label{sec:trace-steps}

To illustrate how the inspectability classification operates in practice, we follow one candidate along the code that actually runs in 2025: \emph{a reply posted by a Blue-verified user, scored for a viewer who does not follow that user}. The \emph{For You} product registers a single top-level pipeline; older pipelines that remain in the repository take no part (App.~\ref{app:bypass-history}).

\begin{enumerate}
    \item \textbf{Sourcing.} The reply is retrieved by \texttt{tweet-mixer}, which merges about 35 out-of-network sources. Which sources run, and how many posts each contributes, is set by parameters (\textbf{Class~II}); several sources query external indexes (\textbf{Class~III}).
    \item \textbf{Pre-scoring filters.} The reply must be less than 48 hours old and must not be a duplicate or a post the viewer has already seen (App.~\ref{app:freshness-filter}). These rules are hard-coded (\textbf{Class~I}).
    \item \textbf{Hydration.} Feature hydrators attach the author's Blue verification status from X's user database, a precomputed similarity score between viewer and post, and, if the Grok layer is enabled, Grok's annotations (Apps.~\ref{app:verification-hydration}, \ref{app:simclusters-user-tweet}, and~\ref{app:grok-hydrator}). All three come from external services (\textbf{Class~III}).
    \item \textbf{Ranking.} The Navi model, an external service, predicts 15 engagement probabilities from a record that includes the author's verification status (App.~\ref{app:verified-model-input}). Their weighted sum, with weights set outside the code, is the base score. As an out-of-network reply, the post is then down-weighted by two rescoring factors, to 0.5625 of that score with the code defaults (\textbf{Class~III} for the model; \textbf{Class~II} for weights and factors).
    \item \textbf{Post-scoring filters.} If enabled, four Grok filters remove posts that Grok flags as gore, adult content, spam, or violent (App.~\ref{app:grok-filters})---visible logic over external labels (\textbf{Class~II} over \textbf{Class~III}).
    \item \textbf{Mixing.} The timeline is assembled from the ranked posts and about twenty other modules, such as advertisements and account recommendations, which are inserted at hard-coded positions (\textbf{Class~I}). A video carousel, however, draws on a product whose code is not in the release (\textbf{Class~IV}, App.~\ref{app:video-product}).
\end{enumerate}

\section{Component Findings: Segmentation, Visibility, Ranking}
\label{sec:findings}

Applying the same traversal to user segmentation, visibility filtering, and ranking surfaced the components in Table~\ref{tab:components}. Summarizing, most components are governed by parameters whose code defaults are neutral or off, so the code shows \textit{which mechanisms exist} but not how strongly they act---we examine three mechanisms in depth in Sec.~\ref{sec:further-investigation}. Furthermore, we can see that the \textit{decisive computations are external}: ranking models, Grok's annotations, and interest-similarity scores come from services that the code only calls. Finally, we note that \textit{code from 2023 persists}: the abuse classifier and the safety labels are carried over unchanged from the 2023 release, whereas the explicit boosts for verified authors in that release no longer run in 2025 (Sec.~\ref{sec:further-investigation}).

\begin{table}[!ht]
\centering
\small
\caption{Components with information that is shown/withheld, and classifications and references (to the appendix).}
\label{tab:components}
\begin{tabularx}{\textwidth}{@{}L{0.13\textwidth} Y L{0.20\textwidth} c L{0.05\textwidth}@{}}
\toprule
\textbf{Component} & \textbf{Information Shown} & \textbf{Information Withheld} & \textbf{Class} & \textbf{Ref.} \\
\midrule
\multicolumn{5}{@{}l}{\textit{User segmentation}} \\
\midrule
Verification tiers & Four tiers (Blue, Gold, Gray, Legacy); Blue status is fetched from the user database and passed to the ranking model & Who sets the Gold, Gray, and Legacy flags & II, IV & \ref{app:verification-tiers}, \ref{app:verification-hydration} \\ \rowsep
Reputation score (TweepCred) & PageRank over the follow graph, scaled to 0--100; verified accounts receive the maximal starting weight; reaches the 2025 timeline only as a model input & Data snapshots and run settings; how much the model uses the score & I, II & \ref{app:tweepcred}, \ref{app:tweepcred-threshold} \\ \rowsep
Interest clusters (SimClusters) & Offline jobs grouping some 20 million accounts into 145,000 communities; similarity scores used for retrieval and ranking & The services that serve the scores; what each cluster means & II, III & \ref{app:simclusters}, \ref{app:simclusters-hydrator} \\
\midrule
\multicolumn{5}{@{}l}{\textit{Visibility filtering}} \\
\midrule
Abuse classifier (2023) & Training code for a multilingual BERT model with eight published enforcement labels & Weights, data, redacted input features; no released pipeline calls it & III & \ref{app:abuse-classifier} \\ \rowsep
Safety labels (2023) & Label names such as \texttt{DoNotAmplify} or \texttt{UntrustedUrl}, carried over unchanged from 2023 & What triggers each label & III & \ref{app:donotamplify} \\ \rowsep
Grok annotations & 13 features per post (App.~Table~\ref{tab:grok-annotations}), read from a cache; they feed four filters, the ranking model's input, and a rescoring factor; off by default & Prompts, model, and the full label set & II, III & \ref{app:grok-hydrator}, \ref{app:grok-filters} \\ \rowsep
Grok topics & 32 category names, used only for metrics and logging & The category IDs (redacted) & III & \ref{app:grok-topics} \\ \rowsep
Freshness filter & A hard-coded 48-hour limit for every scored candidate & Nothing & I & \ref{app:freshness-filter} \\
\midrule
\multicolumn{5}{@{}l}{\textit{Content ranking}} \\
\midrule
Ranking models & Navi (15 heads) scores by default; Phoenix (13 heads) is switched off; screenshots count among the positive engagements & The models and how they use their inputs & III & \ref{app:navi-heads}, \ref{app:phoenix-heads} \\ \rowsep
Ranking weights & 30 weights, all defaulting to 0.0; optional random noise and an external override & Production values & II & \ref{app:model-weights}, \ref{app:weight-noise} \\ \rowsep
Rescoring factors & A chain of multipliers, e.g., 0.75 each for out-of-network posts and for replies & Production values & II & \ref{app:oon-scale} \\
\bottomrule
\end{tabularx}
\end{table}

One further point concerns the Grok labels themselves. Nowhere does the release define what counts as ``slop'', ``spam'', ``low quality'', or a post's political inclination: these judgments are made inside the model, and the code only receives the verdict. This matters because LLMs carry social and political biases~\cite{gallegos2024bias}, which propagate into downstream classification tasks such as hate-speech and misinformation detection~\cite{feng2023pretraining}, and LLM-based content moderation remains error-prone~\cite{kumar2024watch}. Undefined labels produced by such a model can systematically disadvantage some speakers or viewpoints, and the release gives an auditor no way to check whether they do. A label's name is also weak evidence of its criteria; this is demonstrated in Meta's personal AI agent Muse, where blocked URLs are reported to the agent as being ``safety blocklist''---however, the list includes \texttt{nytimes.com}, \texttt{bbc.com}, and \texttt{apnews.com},\footnote{Decoded from files in a filesystem export of a Muse virtual machine from September 2026; see \url{https://beep.acab.taxi/files/muse_vm-2026-09-24T1016.tar.gz}, posted in \url{https://neuromatch.social/@jonny/117334075347648703}; downloaded 26 September 2026.} showing that a label called ``safety'' may carry criteria other than safety. X's Grok labels are visible only by name.

\subsection{Mechanisms Requiring Further Investigation}
\label{sec:further-investigation}

Beyond component-level classification, our traversal surfaced three mechanisms that connect to systemic risks and to further DSA obligations, but whose effect cannot be established from the release. We report them as findings for further investigation and, for each, state what \textit{mechanism} the code shows, what the \emph{implications} are, and which \emph{artifacts} a vetted researcher would have to request through data access under the DSA to resolve it. Each finding thus turns a general systemic-risk hypothesis into a specific question; Sec.~\ref{sec:discussion} returns to how such requests can be structured.

\subsubsection{Political Inclination Labels in the Grok Layer}

\paragraph{Mechanism.} Platforms increasingly embed generative AI in the systems that rank and moderate content, in addition to user-facing assistants.\footnote{P.~Leerssen, \href{https://dsa-observatory.eu/2024/10/16/1864/}{Embedded GenAI on Social Media: Platform Law Meets AI Law}, DSA Observatory, 2024.} X announced this step publicly (App.~\ref{app:tweet-grok-replacement}), and the code shows it: when enabled, the Grok layer attaches 13 annotations to every post in the ranking pipeline (App.~Table~\ref{tab:grok-annotations}). The most sensitive one labels a post's political inclination, with values including left, right, and neutral. Its inline comment states that it is ``used only for metrics tracking'' and ``does not affect the recommendations'', and the release is consistent with that: no filter, rescoring factor, or model input uses it. The label is, however, counted. For each request, the system logs how many left-, right-, and neutral-labeled posts were served, with the viewer's user ID, behind a logging flag that is off by default (App.~\ref{app:tweet-type-predicates}).

\paragraph{Implications.} Inferring the political leaning of posts written by identifiable users comes close to processing special categories of personal data under Art.~9(1) GDPR~\cite{gdpr2016}, and counting labeled posts per viewer permits inferences about the viewer's political exposure and interests. Article~26(3) DSA prohibits presenting advertisements based on profiling that uses such categories. The release supports the claim that the label does not act on ranking, but whether the logged counts feed advertising, profiling, or the training of unreleased models cannot be established from the code.

\paragraph{Artifacts needed.} The production configuration that determines which Grok annotations act on filtering and ranking; the annotation prompts and label definitions, including the political taxonomy, with per-label performance; the production value of the logging flag and the downstream consumers of logs; a sample of stored annotations.

\needspace{6\baselineskip}
\subsubsection{Monetized Amplification}

\paragraph{Mechanism.} The blue checkmark, which on Twitter marked accounts that the platform had verified as authentic and notable, can be bought on X through a Premium subscription. The repository's history shows how the code linking the checkmark to reach has changed (App.~\ref{app:verification-history}). The 2023 releases boosted Blue-verified authors explicitly: they multiplied their scores by 4 for viewers who follow them and by 2 for viewers who do not, let their posts reach non-followers without the usual social proof (a like or follow from the viewer's network), and boosted subscription-eligible creators by 1.1 and 1.3. In the 2025 release, these boosts are gone or sit in code that no longer runs (App.~\ref{app:bypass-history}). What remains is less direct. Verification status is an input to the ranking model, so it can raise or lower scores with a weight that the release does not show (App.~\ref{app:verified-model-input}). A 2023 recommendation service that places Blue-verified posts first by default is still in the release and is still called by the released push-notification service (App.~\ref{app:blue-verified-topk}). The viewer's own Premium status is passed to the configuration system that sets ranking parameters, and the timeline code reserves a category for ``promoted creator'' posts, although the retrieval channel behind it is not wired in. The code thus shows that verification still enters ranking, but not with what weight.

\paragraph{Implications.} The European Commission issued preliminary findings in July 2024 that the checkmark is deceptively designed, since it signals verification but requires only a paid subscription. In December 2025 it fined X EUR~120 million for three breaches, of which the checkmark design was one; the others concerned the advertising repository and researcher data access.\footnote{European Commission, \href{https://ec.europa.eu/commission/presscorner/detail/en/ip_24_3761}{IP/24/3761} (12 July 2024) and \href{https://ec.europa.eu/commission/presscorner/detail/en/ip_25_2934}{IP/25/2934} (5 December 2025).} Neither decision concerned ranking. The code adds that dimension: the explicit 2023 boosts show that the paid badge was wired to reach, and the 2025 model input keeps that link open, which bears on the Art.~34(1)(b) and (c) risks to fundamental rights and civic discourse. The paid badge also raises an open advertising question: whether posts boosted by a flat subscription must be labeled as advertisements under Art.~26(1) DSA, as the European consumer organization BEUC argues.\footnote{C.~Samyn and S.~Rossetti, \href{https://blog.beuc.eu/renting-credibility-how-x-blue-checkmark-became-tool-for-scammers/}{Renting credibility}, BEUC, 29 July 2026.}

\paragraph{Artifacts needed.} How sensitive the ranking model is to verification status (for instance, scores for matched candidates with the feature toggled); the production values of the verified-first ordering and of any parameters keyed on the viewer's Premium status, including their experiment history; the surfaces that the 2023 recommendation service still serves; and the selection criteria of the promoted-creator category.

\subsubsection{Evergreen Content}

\paragraph{Mechanism.} Feeds favor fresh content, and X's ranking pipeline drops every post older than 48 hours before scoring. ``Evergreen'' content is not exempt from this rule; the term refers to content that is selected for additional retrieval routes. Since 2023, an internal table has listed evergreen posts, from which an interest-based index is built (App.~\ref{app:evergreen-label}). The 2025 update adds evergreen retrieval channels for the \emph{For You} timeline: three that use ``evergreen'' user models to find posts, one of them across country borders, and one that retrieves ``evergreen videos'' from the accounts a viewer follows out of a historical index (Apps.~\ref{app:evergreen-factory} and~\ref{app:evergreen-query}). All of them are switched off by default and carry experiment names, and the video channel runs as a test call whose results are discarded. What makes content evergreen is not in the release: neither the producer of the table nor the process that fills the video index. Posts that reach the main ranking pipeline cannot be older than 48 hours, but the \emph{For You} video carousel draws on a product whose code is not released.

\paragraph{Implications.} Deciding that content is evergreen is a decision about which content deserves additional distribution, and its criteria are not in the release. Content selected this way gains retrieval routes that other posts lack, including routes outside a viewer's network and across borders. If these are active, the selection forms a curation channel that users cannot see and outsiders cannot inspect but that would plausibly count among the main parameters that Art.~27(1) DSA requires platforms to set out in their terms and conditions, and it bears on the Art.~34(1)(c) risks to civic discourse.

\paragraph{Artifacts needed.} The criteria and process that select evergreen posts and videos; the production values of the evergreen channels' switches; the configuration of the video product behind the \emph{For You} carousel and the freshness rules that apply there; and a sample of selected items.

\section{Dead-Ends in the Released Code}
\label{sec:dead-ends}

The traversal with its reachability check produces two inventories. The first lists \emph{dead-ends}: components that running code depends on but whose definitions, implementations, or configurations are absent from the release. The second covers references inside code that the 2025 product no longer reaches (Sec.~\ref{sec:orphaned}). The 13 dead-ends we identified cluster around three critical aspects: \textit{model opacity}, since every candidate is scored by external models, and the weights that combine their predictions default to 0.0; \textit{LLM annotation semantics}, since the code shows which signals Grok produces but not how; and \textit{inputs that X maintains outside the release}, such as the evergreen table and the non-Blue verification flags. All three aspects are based on configurations where production values are not public.

For an auditor, each of these aspects raises questions: The code shows that verification status is passed to the ranking model, but not whether it raises a post's score. It shows that Grok's labels feed filters and the model, but not how a post earns the label ``spam'' or ``low quality'', or how often legitimate posts are removed as a result. It shows that some content is selected as ``evergreen'' and given additional retrieval routes, but not by whom or by which criteria. And it shows which engagement types the system predicts, but not which of them it rewards.
Together, these dead-ends foreclose three categories of audit questions: (1)~how much verification status weighs in ranking, (2)~how LLM-based content classification shapes moderation outcomes, and (3)~what engagement behaviors the system is configured to maximize. Table~\ref{tab:dead-ends} lists each dead-end with the missing artifact and the question that access to it would answer.

\begin{table}[!ht]
\centering
\small
\caption{Dead-ends inventory: components that the public code depends on but the release does not contain.}
\label{tab:dead-ends}
\begin{tabularx}{\textwidth}{@{}Y L{0.22\textwidth} L{0.27\textwidth} c c@{}}
\toprule
\textbf{Component} & \textbf{Missing Artifact} & \textbf{What Access Would Unlock} & \textbf{Class} & \textbf{Ref.} \\
\midrule
Navi ranking model & Model, input features, learned weights & How posts are scored; how much verification and Grok labels count & III & \ref{app:navi-heads} \\ \rowsep
Phoenix ranking model & Model; definitions of its engagement types & Whether and how the alternative model ranks & III & \ref{app:phoenix-heads} \\ \rowsep
Interest-similarity scores (SimClusters) & Serving services; switch value & How interest similarity enters ranking & III & \ref{app:simclusters-user-tweet} \\ \rowsep
Reputation score computation (TweepCred) & Job settings, data snapshots & How reputation enters ranking & II & \ref{app:tweepcred} \\ \rowsep
Grok annotation producer & Prompts, model, cache writer & How annotations are generated & III & \ref{app:grok-hydrator} \\ \rowsep
Grok label definitions, incl.\ political inclination & Type definitions & The full label set and political taxonomy & IV & \ref{app:grok-political} \\ \rowsep
Grok topic category IDs & Redacted IDs & Which category each score refers to & III & \ref{app:grok-topics} \\ \rowsep
Evergreen content table & Producer of the table & What makes content ``evergreen'' & IV & \ref{app:evergreen-label} \\ \rowsep
Evergreen video index & Index population & Which videos are retrievable regardless of age & III & \ref{app:evergreen-source} \\ \rowsep
Gold, Gray, and Legacy verification flags & Upstream producer & How the non-Blue tiers are assigned & IV & \ref{app:verification-hydration} \\ \rowsep
Video product behind the \emph{For You} carousel & Entire product implementation & Which retrieval, filters, and ranking apply to videos & IV & \ref{app:video-product} \\ \rowsep
Production values of the 30 ranking weights & Runtime configuration & Which engagement behaviors the system rewards & II & \ref{app:model-weights} \\ \rowsep
On/off switches for the Grok layer, Phoenix, evergreen channels, and logging & Parameter values & Which components run at all & II & \ref{app:grok-filters} \\
\bottomrule
\end{tabularx}
\end{table}

\subsection{Causes and Consequences of Dead-Ends}
\label{sec:dead-ends-method}

Some dead-ends are to be expected. This is driven by two structural forces. First, \textit{trade secrets and the security of the service} bound what any VLOP will disclose: ranking logic, weights, and experiment configurations are treated as confidential, and fully disclosing moderation rules or safety thresholds would let coordinated actors evade them.
Second, \textit{LLM integration introduces a distinct source of opacity}: an LLM annotation layer produces signals through prompts and inference that cannot be reconstructed from code, and its outputs are not reproducible by construction.
These forces explain the dead-ends; they do not make them less consequential. Each dead-end in Table~\ref{tab:dead-ends} is a point at which independent auditing stops, and together they place the most consequential decisions beyond external scrutiny. %
 Whether a given omission is justified by trade secrecy or by the security of the service is itself a question that the release does not allow outsiders to assess. The inventory is therefore best read as a list of questions for regulatory data access.

\subsection{Orphaned Code}
\label{sec:orphaned}
Each update since the first release in March 2023 left earlier code in place, so the release contains code that no identified pipeline reaches. The 2025 commit rebuilt the \emph{For You} product around a single top-level pipeline but left three 2023 pipelines and their components in the repository (App.~\ref{app:bypass-history}). By our count, these files read 36 parameters that are defined nowhere at \texttt{c54bec0}, and every one of them is defined, with its name and code default, in the repository's July 2023 commit. They include the verified-author social-proof bypass (default \texttt{true} in 2023), the positions of recommendation modules (App.~\ref{app:foryou-scorer-mixer}), the switch of a 2023 candidate pipeline (App.~\ref{app:foryou-scorer-candidate}), filters against posts from ``competitor'' accounts and URLs, and the 17 engagement weights of the 2023 ranker. These references look like dead-ends; checked for reachability, they are artifacts of a partial update. The distinction matters for auditors: a release that mixes live and unreachable code without marking which is which invites findings about mechanisms the platform no longer runs, and our own traversal would have produced such findings without the reachability step.

\section{Discussion}
\label{sec:discussion}

Because X's release is partial, the insights it allows about the system are limited; yet they are telling.
A pattern visible across the two releases is a move away from deterministic classifiers and auditable models.
In the 2023 architecture, trust-and-safety decisions were made by supervised classifiers whose architecture was visible even when their weights were not, so a researcher granted access could inspect the model and reason about its decision boundaries. In the 2025 architecture, those decisions are partly driven by Grok annotations. The filters that \emph{apply} these annotations are fully visible, but their \emph{inputs} have shifted to Class~III (Interface only), and audit methods that reason about decision boundaries from model architecture do not straightforwardly extend to this setting.
A second challenge is distinguishing features that \emph{exist} in the pipeline from features that \emph{act} on recommendations. The political-inclination label illustrates this. Within the release it does not act on ranking, but it sits next to annotations that do, its counts are logged per viewer, and unreleased services can read both. Whether a feature acts on recommendations therefore depends on the production system as a whole, and the same ambiguity applies to Class~II parameters throughout the pipeline. Verifying it requires runtime observation in addition to static code analysis. Even so, users are not told that the political leaning of what they see is inferred and recorded at all.
Our dead-ends inventory (Table~\ref{tab:dead-ends}) also points to the minimally required artifacts that would close each gap:

\begin{itemize}
\item \textbf{Model cards for LLM components.} Disclosing Grok's prompt templates, training objectives, and performance metrics per annotation type would move Grok features from Class~III (Interface only) to Class~II (Parameter-dependent), giving auditors a basis to reason about annotation behavior without requiring model weights. 
\item \textbf{Parameter dumps for A/B configurations.} Releasing the production values of parameters such as the 30 ranking weights and the Grok switch would turn Class~II components into Class~I and show which of the existing features act on recommendations.
\item \textbf{A reachability manifest.} Marking which pipelines a product registers, and removing or flagging code that no product reaches, would let auditors separate live dead-ends from artifacts of partial updates without reconstructing the wiring themselves.
\item \textbf{Versioning in step with production.} Beyond closing individual gaps, releases should be versioned in step with production, which would let auditors date their findings and follow how the pipeline changes over time.
\end{itemize}

The Grok integration introduces a new category of provenance failure. Where deterministic classifiers preserve a reproducible mapping from input to output, an LLM annotation layer does not. In X's pipeline, the annotations are generated once per post and cached, so each served decision rests on a fixed label, but an auditor who re-runs the model on the same post may obtain a different one. Releasing Grok's prompts and weights would move the component from Class~III (Interface only) to Class~II (Parameter-dependent), but it would not restore reproducibility. Decision provenance under hybrid recommender--LLM architectures therefore requires more than disclosure.

\subsection{From Mechanisms to Accountability}
\label{sec:accountability}

Code-level insight opens a new avenue for platform auditing when it is paired with the findings of traditional, output-based audits. Such audits have documented risks on X: Twitter's own study found that its algorithm amplified the mainstream political right more than the left in six of seven countries~\cite{huszar2022amplification}; a sock-puppet audit of the 2024 U.S. presidential election found a right-leaning bias in the default timelines of new accounts~\cite{x-audit-2024}; an observational study found that the algorithm amplified low-credibility content~\cite{corsi2024lowcred}; and a randomized experiment found that engagement-based ranking amplified emotionally charged, out-group hostile content~\cite{milli2025engagement}. These studies establish \emph{that} risks arise. Code-level evidence can help show \emph{where} they arise---whether a skew stems from candidate sources, from ranking weights, from paid verification, or from curated channels such as evergreen content.

This distinction matters for accountability and, ultimately, liability. Section~230 of the U.S. Communications Decency Act shields providers from being treated as the publisher of content their users post,\footnote{47 U.S.C. \S~230(c)(1). In \emph{Gonzalez v. Google LLC}, 598 U.S. 617 (2023), the U.S. Supreme Court declined to decide whether the provision covers algorithmic recommendations.} and the DSA likewise exempts hosting services from liability for user content under certain conditions (Art.~6)~\cite{DSA2022}. Both protections attach to content that users post, not to the platform's own design decisions. If an audit can show that a documented risk stemmed from the platform's own mechanisms---a boost for paying accounts, the selection of content for additional reach, or the weights that decide what is ranked first---rather than from user content alone, it produces evidence about how the risk emerged and who shaped it. In the EU, such evidence speaks directly to the obligation of very large platforms to assess and mitigate risks stemming from the design of their recommender systems (Arts.~34 and~35). Producing it at scale will require code-level access that goes beyond X's voluntary release---to code, configuration, and the logs that show which mechanisms are active---through data access under the DSA or comparable regimes.

\section{Limitations and Future Work}
\label{sec:limitations}
Most limitations of this study follow directly from the dead-ends it documents. We can only audit what X has released, and the release omits the components that decide most outcomes: the ranking models, Grok's annotations, and the production configuration. Everything we report is therefore a statement about the released code. We cannot say which code paths are active in production, which parameters take non-default values, how often any mechanism fires, or what effect it has on what users see. Our reachability check is static as well: it follows the pipelines that the release registers, which need not match what X deploys. Complete coverage of a codebase of this scale is also unrealistic; hence, this audit surfaces evidence that particular mechanisms \emph{exist} in the pipeline, not a full account of how they interact. Production is likewise a moving target---in March 2023, X's CEO stated that the algorithm is updated ``every 24 to 48 hours'' (App.~\ref{app:tweet-rapid-updates})---while the public repository has received only two substantive updates since its first release, in July 2023 and September 2025; any snapshot is therefore merely an approximation of production logic. In January 2026, X published a separate, rewritten repository for its recommender system;\footnote{\url{https://github.com/xai-org/x-algorithm}} we did not analyze it, because X's Help Center and transparency materials refer to the repository we audited and our analysis began before the new one was released.
More generally, audits have been criticized for documenting harms already known while leaving the burden of scrutiny with outsiders and the power to act with the audited firm~\cite{devrio2026plaudits}. This applies to our audit as well: it was possible only because X chose to publish its code, hence the releases are under the control of the audited organization itself.
Natural next steps are to pair code-level findings with data access under the DSA and with behavioral methods, such as sock-puppet audits~\cite{TORS}, to confirm which mechanisms are active and measure their downstream effects.

\section{Conclusions}
We have presented a code-level audit methodology: a backwards-walk traversal with a reachability check, a four-tier inspectability classification, and a dead-ends inventory. We applied this methodology to X's 2025 open-source repository, and identified 13 dead-ends on the live serving path as well as 36 parameter references in 2023 code that the 2025 product no longer reaches. The components that decide which posts are shown, and in what order, are Class~III (Interface only) or Class~IV (Opaque): the ranking models run as external services, the Grok annotation layer is interface-only, and the lists that decide evergreen content and verification tiers are maintained outside the code. The release thus does not show what X's recommender does, but it shows where to look, and it turns systemic-risk hypotheses into precise requests for data access under the DSA. The Grok integration indicates that this will become harder: where LLM annotations replace inspectable classifiers, releasing more code does not restore a reproducible mapping from input to decision, and auditors will also need prompts and stored outputs. Our methodology has been designed to remain platform-agnostic and to apply to any recommender system, including hybrid recommender--LLM systems, whose source code is available or can be requested under the DSA.

\section*{Ethical Considerations Statement}
This study analyzes material that X has published: the source code in its public repository and its public documentation. In addition, we quote a small number of documentation and configuration files from a filesystem export of Meta's Muse agent that a third party posted publicly; we use it only to illustrate one point in Sec.~\ref{sec:findings}, quote no personal data, and did not access any Meta system. We did not collect user data, create accounts, interact with X's live service, or involve human participants. The main ethical risk is dual use: describing how the pipeline treats verification, safety labels, or evergreen content could help actors game the system. We mitigate this by describing mechanisms only at the level needed to assess accountability, without optimization guidance, and by discussing nothing that is not already in X's public release. A second risk is misattribution, since code does not show runtime behavior and a release can contain code that no longer runs. We therefore separate live from unreachable code, verify every code claim against a pinned and archived commit, frame mechanisms as hypotheses rather than established effects, and present legal assessments as open questions where the law is unsettled.

\section*{Generative AI Usage Statement}
We used Anthropic's Claude Opus models (versions 4 through 5.5) to analyze the released source code---for instance, to locate definitions and trace references across files---and to help with writing and editing the manuscript. All code claims were verified by the authors against the pinned commit, and all text was reviewed and revised by the authors, who take full responsibility for the content.

\newpage

\appendix

\section{Verified X Accounts and Posts Referenced}
\label{app:xaccounts}
\begin{itemize}
    \item \texttt{@X} --- Official platform account for corporate announcements.
    \item \texttt{@XEng} --- Engineering team; publishes system updates and code release notes.
    \item \texttt{@XDevelopers} --- Developer relations and API announcements.
    \item \texttt{@elonmusk} --- Owner; platform-level policy and feature updates.
    \item \texttt{@nikitabier} --- Former Head of product @X; product-level and engineering insights.
\end{itemize}

\section{X Posts Referenced}
\label{app:xposts}

\label{app:tweet-opensource}
\noindent\textit{Elon Musk (@elonmusk)}:
``By going open source, any interested person (including you) will be able to determine, with reasonable accuracy, what will probably be shown to users, independent of our posts on this platform.''
\url{https://x.com/elonmusk/status/1641874582473695246}
(Date: 31 March 2023)

\medskip
\label{app:tweet-rapid-updates}
\noindent\textit{Elon Musk (@elonmusk)}:
Musk stated that, following the initial open-source release, X's recommendation algorithm would be updated ``every 24 to 48 hours,'' with substantial improvements to be reflected in the open-source repository.
\url{https://x.com/elonmusk/status/1641906265944203267}
(Date: 31 March 2023)

\medskip
\label{app:tweet-dec2024}
\noindent\textit{Elon Musk (@elonmusk)}:
Musk announced a major overhaul of the recommendation system, reporting a $10\times$ increase in the number of parameters in the sparse model and a $3\times$ increase in training compute relative to the prior iteration.
\url{https://x.com/elonmusk/status/1869527728904642591}
(Date: 19 December 2024)

\medskip
\label{app:tweet-grok-replacement}
\noindent\textit{Elon Musk (@elonmusk)}:
Musk stated that X's recommendation algorithm would be replaced with a ``lightweight version of Grok,'' indicating a further expansion of LLM-generated signals in the recommendation pipeline.
\url{https://x.com/elonmusk/status/1918592668307010019}
(Date: 03 May 2025)

\medskip
\noindent\textit{Elon Musk (@elonmusk)} ---
Musk publicly invited users to reply to a post with ``divisive facts'' for \texttt{@Grok} training, describing these as politically incorrect but factually true.
\url{https://x.com/elonmusk/status/1936493967320953090}
(Date: 21 June 2025)

\medskip
\label{app:grok-future}
\noindent\textit{Elon Musk (@elonmusk)}:
Summary: Musk reports rapid ongoing changes to X's recommendation system, including plans to open-source the full codebase, roll out a new Grok model powering a ``fully completed'' algorithm, and continued improvements in posting activity, feed quality, and AI-based features.
\url{https://x.com/elonmusk/status/1996989600602996811}
(Date: 05 December 2025)

\section{Source Code References}
\label{app:source-code-refs}

All source code references in this paper are pinned to commit
\texttt{c54bec0} (full SHA:
\texttt{c54bec0d\br{}4e029fe3\br{}4926ef32\br{}58a86cca\br{}cc0d0182})
of the repository at \url{https://github.com/twitter/the-algorithm}.
To ensure long-term availability, both the repository and this specific
revision have been permanently archived via Software
Heritage~\cite{Cosmo_2023} and can be accessed using the following
persistent identifier (SWHID):

\begin{center}
\small\texttt{swh:1:rev:c54bec0d4e029fe34926ef3258a86ccacc0d0182}
\end{center}

\noindent The archived revision is available at:
\url{https://archive.softwareheritage.org/swh:1:rev:c54bec0d4e029fe34926ef3258a86ccacc0d0182;origin=https://github.com/twitter/the-algorithm;visit=swh:1:snp:84495b2cac570e5c36cab944ecab3bd045e00223}

\subsection{Verification and Segmentation}

\begin{appitems}
  \item \label{app:verification-tiers} \textit{Verification tier feature definitions} ---
  \texttt{AuthorIs\br{}Blue\br{}Verified\br{}Feature}, \texttt{AuthorIs\br{}Gold\br{}Verified\br{}Feature}, \texttt{AuthorIs\br{}Gray\br{}Verified\br{}Feature}, \texttt{AuthorIs\br{}Legacy\br{}Verified\br{}Feature}, \texttt{AuthorIsCreatorFeature}.\\
  \url{https://github.com/twitter/the-algorithm/blob/c54bec0/home-mixer/server/src/main/scala/com/twitter/home_mixer/model/HomeFeatures.scala#L74-L78}

  \item \label{app:verification-hydration} \textit{Verification feature hydration from Gizmoduck user service} ---
  Blue verification extracted from \texttt{user.safety\br{}.flatMap\br{}(\_.isBlue\br{}Verified)} (line~115) and written to \texttt{AuthorIs\br{}Blue\br{}Verified\br{}Feature} (line~199). The verified-organization fields are read (lines~117--121) but feed only the author data record, not the Gold or Gray features; Gold, Gray, and Legacy are copied from upstream service responses.\\
  \url{https://github.com/twitter/the-algorithm/blob/c54bec0/home-mixer/server/src/main/scala/com/twitter/home_mixer/functional_component/feature_hydrator/GizmoduckAuthorFeatureHydrator.scala#L115-L199}

  \item \label{app:social-context-filter} \textit{Social context filter with verified-author bypass} ---
  \texttt{Enable\br{}Verified\br{}Author\br{}Social\br{}Context\br{}Bypass\br{}Param} consumed at line~31; \texttt{isVerifiedAuthor()} aggregates all four tiers at lines~53--57.\\
  \url{https://github.com/twitter/the-algorithm/blob/c54bec0/home-mixer/server/src/main/scala/com/twitter/home_mixer/product/for_you/filter/SocialContextFilter.scala#L26-L57}

  \item \label{app:creator-multipliers} \textit{Creator in-network and out-of-network multiplier parameters} ---
  \texttt{Creator\br{}InNetwork\br{}Multiplier\br{}Param} (default 1.0, range 0.0--100.0) and \texttt{Creator\br{}OutOfNetwork\br{}Multiplier\br{}Param} (default 1.0, range 0.0--100.0).\\
  \url{https://github.com/twitter/the-algorithm/blob/c54bec0/home-mixer/server/src/main/scala/com/twitter/home_mixer/product/scored_tweets/param/ScoredTweetsParam.scala#L447-L461}

  \item \label{app:verification-predicates} \textit{Verification tier predicates for metrics} ---
  \texttt{author\br{}\_is\br{}\_blue\br{}\_verified}, \texttt{author\br{}\_is\br{}\_gold\br{}\_verified}, \texttt{author\br{}\_is\br{}\_gray\br{}\_verified}, \texttt{author\br{}\_is\br{}\_legacy\br{}\_verified}.\\
  \url{https://github.com/twitter/the-algorithm/blob/c54bec0/home-mixer/server/src/main/scala/com/twitter/home_mixer/functional_component/decorator/builder/HomeTweetTypePredicates.scala#L251-L254}
\end{appitems}

\subsection{TweepCred Reputation}

\begin{appitems}
  \item \label{app:tweepcred} \textit{TweepCred reputation scaling formula} ---
  Converts raw PageRank to a 0--100 scale: $\text{round}(130 + 5.21 \cdot \ln(\text{raw}))$, clamped to $[0, 100]$; a comment says the constants map maximum and minimum PageRank to about 95 and 15. Post-adjustment penalizes high following-to-follower ratios. The PageRank prior in \texttt{UserMass} gives accounts flagged as verified the maximum mass (lines~40--41).\\
  \url{https://github.com/twitter/the-algorithm/blob/c54bec0/src/scala/com/twitter/graph/batch/job/tweepcred/Reputation.scala#L12-L49}\\
  \url{https://github.com/twitter/the-algorithm/blob/c54bec0/src/scala/com/twitter/graph/batch/job/tweepcred/UserMass.scala#L28-L62}

  \item \label{app:tweepcred-threshold} \textit{TweepCred threshold in follow-recommendations service} ---
  \texttt{TweepCredThreshold = 40} (line~38) selects the viewer's recent follows with TweepCred below 40 (line~24) as seed edges for follow recommendations.\\
  \url{https://github.com/twitter/the-algorithm/blob/c54bec0/follow-recommendations-service/common/src/main/scala/com/twitter/follow_recommendations/common/stores/LowTweepCredFollowStore.scala#L14-L38}

  \item \label{app:tweepcred-search} \textit{TweepCred filter in search (Earlybird)} ---
  \texttt{minTweepCredFilter} applied via \texttt{BadUserRep\br{}Filter.\br{}getBadUserRep\br{}Filter()}.\\
  \url{https://github.com/twitter/the-algorithm/blob/c54bec0/src/java/com/twitter/search/earlybird/search/queries/BadUserRepFilter.java#L29}
\end{appitems}

\subsection{SimClusters and Communities}

\begin{appitems}
  \item \label{app:simclusters} \textit{SimClusters KnownFor data sources} ---
  Two KnownFor datasets (Dec11, Updated) covering $\sim$20M producers across $\sim$145K clusters; the three model versions (Dec11, Updated, 2020) are defined in \texttt{ModelVersions}.\\
  \url{https://github.com/twitter/the-algorithm/blob/c54bec0/src/scala/com/twitter/simclusters_v2/scalding/KnownForSources.scala#L77-L87}\\
  \url{https://github.com/twitter/the-algorithm/blob/c54bec0/src/scala/com/twitter/simclusters_v2/common/ModelVersions.scala#L11-L13}

  \item \label{app:simclusters-hydrator} \textit{SimClusters engagement similarity hydrator} ---
  Fetches precomputed viewer--tweet engagement-similarity scores from an external service, gated by a decider.\\
  \url{https://github.com/twitter/the-algorithm/blob/c54bec0/home-mixer/server/src/main/scala/com/twitter/home_mixer/functional_component/feature_hydrator/SimClustersEngagementSimilarityFeatureHydrator.scala}

  \item \label{app:simclusters-user-tweet} \textit{SimClusters user--tweet scores hydrator} ---
  Fetches a precomputed viewer--tweet SimClusters (20M\_145K\_2020) dot product from an external Strato column (line~35), only for posts with at least nine likes (line~54).\\
  \url{https://github.com/twitter/the-algorithm/blob/c54bec0/home-mixer/server/src/main/scala/com/twitter/home_mixer/functional_component/feature_hydrator/SimClustersUserTweetScoresHydrator.scala#L35-L54}
\end{appitems}

\subsection{Grok Integration}

\begin{appitems}
  \item \label{app:grok-hydrator} \textit{Grok annotations feature hydrator} ---
  \texttt{BulkCandidate\br{}Feature\br{}Hydrator} that attaches 13 Grok-derived features during the hydration phase, read from a per-post cache (\texttt{GrokPostAnnotations}); runs only when \texttt{EnableGrokAnnotations} is on (default \texttt{false}, \texttt{HomeGlobalParams} line~1367).\\
  \url{https://github.com/twitter/the-algorithm/blob/c54bec0/home-mixer/server/src/main/scala/com/twitter/home_mixer/functional_component/feature_hydrator/GrokAnnotationsFeatureHydrator.scala#L46-L76}

  \item \label{app:grok-features} \textit{Grok feature definitions} ---
  All Grok features defined as \texttt{Feature[TweetCandidate, ...]} objects: safety (gore, NSFW, spam, violent), quality (low quality, slop score), content (tags, top category, category data record, OCR, sunny score), political inclination, and the raw annotations.\\
  \url{https://github.com/twitter/the-algorithm/blob/c54bec0/home-mixer/server/src/main/scala/com/twitter/home_mixer/model/HomeFeatures.scala#L138-L225}\\
  \url{https://github.com/twitter/the-algorithm/blob/c54bec0/home-mixer/server/src/main/scala/com/twitter/home_mixer/model/HomeFeatures.scala#L468-L470}

  \item \label{app:grok-political} \textit{GrokPoliticalInclinationFeature with ``metrics only'' comment} ---
  Line~221: ``// Used only for metrics tracking. Does not affect the recommendations.''\\
  \url{https://github.com/twitter/the-algorithm/blob/c54bec0/home-mixer/server/src/main/scala/com/twitter/home_mixer/model/HomeFeatures.scala#L221-L223}

  \item \label{app:grok-topics} \textit{Grok topic category mapping} ---
  32 top-level categories (Sports, News, Business \& Finance, Cryptocurrency, Technology, Science, Gaming, etc.).\\
  \url{https://github.com/twitter/the-algorithm/blob/c54bec0/home-mixer/server/src/main/scala/com/twitter/home_mixer/model/GrokTopics.scala#L3-L38}

  \item \label{app:grok-filters} \textit{Grok-based visibility filters} ---
  \texttt{GrokGoreFilter}, \texttt{GrokNsfwFilter}, \texttt{GrokSpamFilter}, and \texttt{GrokViolentFilter} read the composite \texttt{GrokAnnotationsFeature} as post-scoring filters; each is gated by a feature switch that defaults to \texttt{false} (\texttt{HomeGlobalParams} lines~1237--1264).\\
  \url{https://github.com/twitter/the-algorithm/blob/c54bec0/home-mixer/server/src/main/scala/com/twitter/home_mixer/functional_component/filter/GrokSpamFilter.scala}\\
  \url{https://github.com/twitter/the-algorithm/blob/c54bec0/home-mixer/server/src/main/scala/com/twitter/home_mixer/functional_component/filter/GrokViolentFilter.scala}
\end{appitems}

\begin{table}[!ht]
\centering
\small
\caption{Grok annotation features exposed in the recommendation pipeline (App.~\ref{app:grok-features}).}
\label{tab:grok-annotations}
\begin{tabular}{ll}
\toprule
\textbf{Safety \& Quality} & \textbf{Content Characterization} \\
\midrule
\texttt{GrokIsGoreFeature} & \texttt{GrokTopCategoryFeature} \\
\texttt{GrokIsNsfwFeature} & \texttt{GrokCategoryDataRecordFeature} \\
\texttt{GrokIsSpamFeature} & \texttt{GrokTagsFeature} \\
\texttt{GrokIsViolentFeature} & \texttt{GrokIsOcrFeature} \\
\texttt{GrokIsLowQualityFeature} & \texttt{GrokSunnyScoreFeature} \\
\texttt{GrokSlopScoreFeature} & \texttt{Grok\br{}Political\br{}Inclination\br{}F.} \\
\texttt{GrokAnnotationsFeature} & \\
\bottomrule
\end{tabular}
\end{table}

\subsection{Trust and Safety Models}

\begin{appitems}
  \item \label{app:abuse-classifier} \textit{Abusive content classifier (multilingual BERT)} ---
  Eight-label multi-label head with the labels disclosed (lines~37--46); the continuous input-feature names (\texttt{cat\_names}, lines~17--21) and data queries are redacted; AdamW (line~70), PR-AUC (line~135).\\
  \url{https://github.com/twitter/the-algorithm/blob/c54bec0/trust_and_safety_models/abusive/abusive_model.py#L17-L135}

  \item \label{app:trust-safety-readme} \textit{Trust and safety models README} ---
  Documents \texttt{pToxicity} (borderline content) and \texttt{pAbuse} (TOS-violating content) distinction.\\
  \url{https://github.com/twitter/the-algorithm/blob/c54bec0/trust_and_safety_models/README.md}

  \item \label{app:donotamplify} \textit{DoNotAmplify visibility rules} ---
  \texttt{DoNotAmplifyDropRule} (line~619) and \texttt{DoNotAmplifyAllViewersDropRule} (line~622) in tweet label rules.\\
  \url{https://github.com/twitter/the-algorithm/blob/c54bec0/visibilitylib/src/main/scala/com/twitter/visibility/rules/TweetLabelRules.scala#L619-L623}
\end{appitems}

\subsection{Content Ranking and Engagement}

\begin{appitems}
  \item \label{app:phoenix-heads} \textit{Phoenix predicted score features (13 engagement heads)} ---
  One sealed trait with 13 heads (lines~174--189): fav, reply, retweet, two good-click variants, good profile click, video quality view, share, dwell, open link, screenshot (lines~138--146; it reuses \texttt{OpenLinkParam}), bookmark, and one negative head. Phoenix scoring is gated by \texttt{EnablePhoenixScorerParam} (default \texttt{false}).\\
  \url{https://github.com/twitter/the-algorithm/blob/c54bec0/home-mixer/server/src/main/scala/com/twitter/home_mixer/model/PhoenixPredictedScoreFeature.scala#L16-L192}

  \item \label{app:model-weights} \textit{Model weights (feature-switch-bounded parameters)} ---
  30 weight parameters, all defaulting to 0.0: 22 in $[-10{,}000, 10{,}000]$, \texttt{ReportParam} in $[-20{,}000, 0]$, two in $[-1{,}000, 0]$, and five dwell weights in $[0, 1{,}000]$.\\
  \url{https://github.com/twitter/the-algorithm/blob/c54bec0/home-mixer/server/src/main/scala/com/twitter/home_mixer/param/HomeGlobalParams.scala#L786-L1028}

  \item \label{app:weight-noise} \textit{Heavy ranker weight hydrator with Beta-distributed noise} ---
  Optionally multiplies each weight by $(1 + b)$ with $b$ drawn from one global Beta($\alpha$, $\beta$), seeded per user and day; both switches default to \texttt{false}.\\
  \url{https://github.com/twitter/the-algorithm/blob/c54bec0/home-mixer/server/src/main/scala/com/twitter/home_mixer/functional_component/feature_hydrator/HeavyRankerWeightsQueryFeatureHydrator.scala#L26-L77}

  \item \label{app:oon-scale} \textit{Out-of-network scale factor} ---
  \texttt{OutOfNetwork\br{}ScaleFactor\br{}Param} with default 0.75.\\
  \url{https://github.com/twitter/the-algorithm/blob/c54bec0/home-mixer/server/src/main/scala/com/twitter/home_mixer/product/scored_tweets/param/ScoredTweetsParam.scala#L463-L469}

  \item \label{app:screenshot-action} \textit{Screenshot engagement action definition} ---
  \texttt{ClientTweetTakeScreenshot = 1091} in the unified user actions thrift; Phoenix's \texttt{CLIENT\br{}\_\br{}TWEET\br{}\_\br{}TAKE\br{}\_\br{}SCREENSHOT} comes from an unreleased enum.\\
  \url{https://github.com/twitter/the-algorithm/blob/c54bec0/unified_user_actions/thrift/src/main/thrift/com/twitter/unified_user_actions/action_info.thrift#L373-L375}
\end{appitems}

\subsection{Freshness and Evergreen}

\begin{appitems}
  \item \label{app:freshness-filter} \textit{48-hour freshness cutoff} ---
  \texttt{Custom\br{}Snowflake\br{}Id\br{}Age\br{}Filter(StaticParam(48.hours))}, the first of the scored-tweets pipeline's global filters (line~473), applied to every candidate.\\
  \url{https://github.com/twitter/the-algorithm/blob/c54bec0/home-mixer/server/src/main/scala/com/twitter/home_mixer/product/scored_tweets/ScoredTweetsRecommendationPipelineConfig.scala#L473}

  \item \label{app:age-filter-impl} \textit{Custom\br{}Snowflake\br{}Id\br{}Age\br{}Filter implementation} ---
  Derives content age from Snowflake IDs; keeps candidates where \texttt{queryTime.since(creationTime) <= maxAge}.\\
  \url{https://github.com/twitter/the-algorithm/blob/c54bec0/home-mixer/server/src/main/scala/com/twitter/home_mixer/product/scored_tweets/filter/CustomSnowflakeIdAgeFilter.scala#L24-L49}

  \item \label{app:evergreen-factory} \textit{Evergreen videos candidate pipeline factory} ---
  Builds candidate pipeline for retrieving evergreen-labeled videos from dedicated index.\\
  \url{https://github.com/twitter/the-algorithm/blob/c54bec0/tweet-mixer/server/src/main/scala/com/twitter/tweet_mixer/candidate_pipeline/EvergreenVideosCandidatePipelineConfigFactory.scala#L26-L43}

  \item \label{app:evergreen-source} \textit{Historical evergreen videos candidate source} ---
  Fetches evergreen videos via Strato's \texttt{Historical\br{}Search\br{}By\br{}User\br{}Ids\br{}Client\br{}Column}.\\
  \url{https://github.com/twitter/the-algorithm/blob/c54bec0/tweet-mixer/server/src/main/scala/com/twitter/tweet_mixer/candidate_source/evergreen_videos/HistoricalEvergreenVideosCandidateSource.scala#L15-L26}
\end{appitems}

\subsection{\emph{For You} Timeline Configuration Files}

\begin{appitems}
  \item \label{app:foryouparam} \textit{ForYouParam definition file} ---
  Contains 702 lines of parameter definitions; the 2023 parameters read by the orphaned mixers (Sec.~\ref{sec:orphaned}) are absent from it.\\
  \url{https://github.com/twitter/the-algorithm/blob/c54bec0/home-mixer/server/src/main/scala/com/twitter/home_mixer/product/for_you/param/ForYouParam.scala}

  \item \label{app:foryou-scorer-mixer} \textit{ForYou\br{}Timeline\br{}Scorer\br{}Mixer\br{}Pipeline\br{}Config} ---
  Orphaned 2023 code (Sec.~\ref{sec:orphaned}); reads five parameters that are undefined at \texttt{c54bec0}, controlling cache behavior and module placement: \texttt{TweetPreviewsPositionParam}, \texttt{WhoToFollow\br{}Position\br{}Param}, \texttt{WhoToSubscribePositionParam}, and \texttt{FlipInlineInjectionModulePosition} (lines~268--283), \texttt{ClearCacheOnPtr} (lines~363--364). This mixer is not registered in the 2025 \emph{For You} product.\\
  \url{https://github.com/twitter/the-algorithm/blob/c54bec0/home-mixer/server/src/main/scala/com/twitter/home_mixer/product/for_you/ForYouTimelineScorerMixerPipelineConfig.scala#L266-L284}

  \item \label{app:foryou-scorer-candidate} \textit{ForYouTimelineScorerCandidatePipelineConfig} ---
  Gates the timeline scorer candidate pipeline with \texttt{EnableTimelineScorerCandidatePipelineParam} (lines~97--98); \texttt{SocialContextFilter} runs in this pipeline (line~196).\\
  \url{https://github.com/twitter/the-algorithm/blob/c54bec0/home-mixer/server/src/main/scala/com/twitter/home_mixer/product/for_you/ForYouTimelineScorerCandidatePipelineConfig.scala#L97-L98}

  \item \label{app:served-candidate-kafka} \textit{ServedCandidateKeysKafkaSideEffect} ---
  Gates Kafka publishing of served-candidate keys with \texttt{EnableServed\br{}Candidate\br{}Kafka\br{}Publishing\br{}Param} (line~53).\\
  \url{https://github.com/twitter/the-algorithm/blob/c54bec0/home-mixer/server/src/main/scala/com/twitter/home_mixer/product/for_you/side_effect/ServedCandidateKeysKafkaSideEffect.scala#L53}
\end{appitems}

\subsection{Mechanisms Requiring Further Investigation}

\begin{appitems}
  \item \label{app:tweet-type-predicates} \textit{Political-inclination metrics predicates and their logging} ---
  Lines~389--400 define \texttt{grok\_politics\_neutral}, \texttt{\_left}, and \texttt{\_right}, each commented ``Purely for metrics tracking. Does not affect the recommendations.'' \texttt{ClientEventsBuilder} emits one ``served tweets'' event per predicate with a per-request count (lines~117--126); the client-event side effect logs it with the viewer's user ID. The full annotation, including the label, is written to the per-viewer scored-tweets cache (\texttt{CachedScoredTweetsSideEffect}, line~154).\\
  \url{https://github.com/twitter/the-algorithm/blob/c54bec0/home-mixer/server/src/main/scala/com/twitter/home_mixer/functional_component/decorator/builder/HomeTweetTypePredicates.scala#L389-L400}\\
  \url{https://github.com/twitter/the-algorithm/blob/c54bec0/home-mixer/server/src/main/scala/com/twitter/home_mixer/functional_component/side_effect/ClientEventsBuilder.scala#L117-L126}\\
  \url{https://github.com/twitter/the-algorithm/blob/c54bec0/home-mixer/server/src/main/scala/com/twitter/home_mixer/product/scored_tweets/side_effect/CachedScoredTweetsSideEffect.scala#L154}

  \item \label{app:blue-verified-topk} \textit{Blue-verified top-K ordering in cr-mixer} ---
  When \texttt{EnableBlueVerifiedTopK} (feature switch \texttt{twistly\_core\_blue\br{}\_verified\br{}\_top\_k}, default \texttt{true}) is on, Blue-verified candidates are placed ahead of all others (lines~133--145). Unchanged since the 2023 release.\\
  \url{https://github.com/twitter/the-algorithm/blob/c54bec0/cr-mixer/server/src/main/scala/com/twitter/cr_mixer/candidate_generation/CrCandidateGenerator.scala#L133-L145}\\
  \url{https://github.com/twitter/the-algorithm/blob/c54bec0/cr-mixer/server/src/main/scala/com/twitter/cr_mixer/param/RankerParams.scala#L23-L27}

  \item \label{app:evergreen-query} \textit{Evergreen channels, query, and freshness parameters} ---
  The evergreen video query contains only the viewer's followed accounts (plus the viewer) and a page size, with no age bound (lines~20--30). Enable flags, all defaulting to \texttt{false}: \texttt{ContentExplorationEvergreenDRTweetTweetEnabled} (line~679), \texttt{EvergreenDRUserTweetEnabled} (line~839), \texttt{EvergreenDRCrossBorderUserTweetEnabled} (line~865), \texttt{EvergreenVideosEnabled} (line~1335); \texttt{MaxTweetAgeHoursParam} defaults to 24~h, bounded at 1--720~h (lines~40--46). The dark-traffic call is \texttt{EvergreenVideosSideEffect} (``calls evergreen\_videos dark traffic'', line~19), gated by a decider (line~383 of the params file). The channels are wired into the home product at lines~317--328 of \texttt{HomeRecommendedTweetsRecommendationPipelineConfig}; \texttt{home-mixer} caps evergreen candidates via \texttt{EnableEvergreenDeepRetrievalMaxCountParam} (default \texttt{false}, line~675 of \texttt{ScoredTweetsParam}).\\
  \url{https://github.com/twitter/the-algorithm/blob/c54bec0/tweet-mixer/server/src/main/scala/com/twitter/tweet_mixer/functional_component/transformer/EvergreenVideosQueryTransformer.scala#L20-L30}\\
  \url{https://github.com/twitter/the-algorithm/blob/c54bec0/tweet-mixer/server/src/main/scala/com/twitter/tweet_mixer/param/TweetMixerGlobalParams.scala#L40-L46}\\
  \url{https://github.com/twitter/the-algorithm/blob/c54bec0/tweet-mixer/server/src/main/scala/com/twitter/tweet_mixer/param/TweetMixerGlobalParams.scala#L1335-L1339}\\
  \url{https://github.com/twitter/the-algorithm/blob/c54bec0/tweet-mixer/server/src/main/scala/com/twitter/tweet_mixer/functional_component/side_effect/EvergreenVideosSideEffect.scala#L19}\\
  \url{https://github.com/twitter/the-algorithm/blob/c54bec0/tweet-mixer/server/src/main/scala/com/twitter/tweet_mixer/product/home_recommended_tweets/HomeRecommendedTweetsRecommendationPipelineConfig.scala#L317-L328}

  \item \label{app:evergreen-label} \textit{Evergreen content label} ---
  Evergreen post IDs are read from the internal BigQuery table \texttt{twttr-\br{}recos-\br{}ml-\br{}prod.\br{}simclusters.\br{}evergreen\_\br{}content\_\br{}data} (lines~30--47); the table's producer is not in the release. The SimClusters evergreen-content index built from it caps post age at 48~hours (line~599). Both date from the 2023 release.\\
  \url{https://github.com/twitter/the-algorithm/blob/c54bec0/src/scala/com/twitter/simclusters_v2/scio/bq_generation/sql/evergreen_content_user_tweet_action_pair_generation.sql#L30-L47}\\
  \url{https://github.com/twitter/the-algorithm/blob/c54bec0/src/scala/com/twitter/simclusters_v2/scio/bq_generation/simclusters_index_generation/EngagementEventBasedClusterToTweetIndexGenerationJob.scala#L590-L599}

  \item \label{app:bypass-history} \textit{Verified-author bypass: 2023 definition and 2025 wiring} ---
  In July 2023, \texttt{EnableVerified\br{}Author\br{}Social\br{}Context\br{}Bypass\br{}Param} was defined as \texttt{for\_you\_\br{}enable\_\br{}verified\_\br{}author\_\br{}social\_\br{}context\_\br{}bypass} with default \texttt{true} (lines~23--27), alongside \texttt{EnableTimelineScorerCandidatePipelineParam} (default \texttt{false}, lines~29--33). In 2025 the \emph{For You} product registers only \texttt{ForYouMixerPipelineConfig} (lines~97--100), so the mixer that contains \texttt{SocialContextFilter} is unreachable.\\
  \url{https://github.com/twitter/the-algorithm/blob/72eda9a/home-mixer/server/src/main/scala/com/twitter/home_mixer/product/for_you/param/ForYouParam.scala#L17-L33}\\
  \url{https://github.com/twitter/the-algorithm/blob/c54bec0/home-mixer/server/src/main/scala/com/twitter/home_mixer/product/for_you/ForYouProductPipelineConfig.scala#L97-L100}

  \item \label{app:verification-history} \textit{Verification and creator boosts in the 2023 releases} ---
  March 2023: \texttt{VerifiedAuthorScalingScorer} applied \texttt{BlueVerified\br{}Author\br{}InNetwork\br{}Multiplier\br{}Param} (default 4.0) and \texttt{BlueVerified\br{}Author\br{}OutOfNetwork\br{}Multiplier\br{}Param} (default 2.0). July 2023: \texttt{RescoreCreators} boosted ``authors who are eligible subscription content creators'' by \texttt{CreatorInNetworkMultiplierParam} (default 1.1) and \texttt{CreatorOutOfNetworkMultiplierParam} (default 1.3). None of these scorers exists at \texttt{c54bec0}.\\
  \url{https://github.com/twitter/the-algorithm/blob/ef4c5eb/home-mixer/server/src/main/scala/com/twitter/home_mixer/param/HomeGlobalParams.scala#L89-L101}\\
  \url{https://github.com/twitter/the-algorithm/blob/ef4c5eb/home-mixer/server/src/main/scala/com/twitter/home_mixer/functional_component/scorer/VerifiedAuthorScalingScorer.scala#L55-L56}\\
  \url{https://github.com/twitter/the-algorithm/blob/72eda9a/home-mixer/server/src/main/scala/com/twitter/home_mixer/product/scored_tweets/scorer/RescoringFactorProvider.scala#L37-L53}

  \item \label{app:verified-model-input} \textit{Blue verification as a heavy-ranker input} ---
  \texttt{GizmoduckFeaturesAdapter} writes \texttt{AUTHOR\_IS\_BLUE\_VERIFIED} into a data record (lines~13, 27), which \texttt{GizmoduckAuthorFeatureHydrator} attaches to each candidate (line~202); the hydrator is wired unconditionally into the scoring pipeline (line~211), and \texttt{NaviModelScorer} converts all data-record features into the model request (lines~63--64).\\
  \url{https://github.com/twitter/the-algorithm/blob/c54bec0/home-mixer/server/src/main/scala/com/twitter/home_mixer/functional_component/feature_hydrator/adapters/gizmoduck_features/GizmoduckFeaturesAdapter.scala#L13-L29}\\
  \url{https://github.com/twitter/the-algorithm/blob/c54bec0/home-mixer/server/src/main/scala/com/twitter/home_mixer/functional_component/feature_hydrator/GizmoduckAuthorFeatureHydrator.scala#L193-L202}\\
  \url{https://github.com/twitter/the-algorithm/blob/c54bec0/home-mixer/server/src/main/scala/com/twitter/home_mixer/product/scored_tweets/scoring_pipeline/ScoredTweetsModelScoringPipelineConfig.scala#L211}\\
  \url{https://github.com/twitter/the-algorithm/blob/c54bec0/home-mixer/server/src/main/scala/com/twitter/home_mixer/functional_component/scorer/NaviModelScorer.scala#L63-L64}

  \item \label{app:navi-heads} \textit{Navi heavy-ranker heads} ---
  The 15 prediction heads whose weighted sum forms the default score; Navi runs unconditionally, while Phoenix is gated (line~311 of the scoring pipeline).\\
  \url{https://github.com/twitter/the-algorithm/blob/c54bec0/home-mixer/server/src/main/scala/com/twitter/home_mixer/model/PredictedScoreFeature.scala#L294-L312}\\
  \url{https://github.com/twitter/the-algorithm/blob/c54bec0/home-mixer/server/src/main/scala/com/twitter/home_mixer/product/scored_tweets/scoring_pipeline/ScoredTweetsModelScoringPipelineConfig.scala#L309-L312}

  \item \label{app:video-product} \textit{Video product behind the \emph{For You} carousel} ---
  The \emph{For You} mixer includes \texttt{forYouScoredVideo\br{}Tweets\br{}Candidate\br{}Pipeline\br{}Config} (line~262); the product registry instantiates \texttt{ScoredVideo\br{}Tweets\br{}Product\br{}Pipeline\br{}Config} (lines~13, 38--39), a class that does not exist in the release.\\
  \url{https://github.com/twitter/the-algorithm/blob/c54bec0/home-mixer/server/src/main/scala/com/twitter/home_mixer/product/for_you/ForYouMixerPipelineConfig.scala#L262}\\
  \url{https://github.com/twitter/the-algorithm/blob/c54bec0/home-mixer/server/src/main/scala/com/twitter/home_mixer/product/HomeProductPipelineRegistryConfig.scala#L13-L39}
\end{appitems}

\bibliographystyle{ACM-Reference-Format}
\bibliography{references}

\end{document}